%% file: main.tex
\documentclass[aps,prd,superscriptaddress,nofootinbib,floatfix,preprint]{revtex4}

\usepackage{graphicx}   

\usepackage{comment}

\newcommand{\beq}{\begin{equation}}
\newcommand{\eeq}{\end{equation}}
\newcommand{\beqa}{\begin{eqnarray}}
\newcommand{\eeqa}{\end{eqnarray}}
\newcommand{\bpr}{\begin{problem}}
\newcommand{\epr}{\end{problem}}
\newcommand{\bcent}{\begin{center}}
\newcommand{\ecent}{\end{center}}
\newcommand{\bfig}{\begin{figure}}
\newcommand{\efig}{\end{figure}}
\newcommand{\bpc}{\begin{picture}}
\newcommand{\epc}{\end{picture}}
\newcommand{\barr}{\begin{array}}
\newcommand{\earr}{\end{array}}
\newcommand{\bitm}{\begin{itemize}}
\newcommand{\eitm}{\end{itemize}}
\newcommand{\bright}{\begin{flushright}}
\newcommand{\eright}{\end{flushright}}
\newcommand{\bminip}{\begin{minipage}}
\newcommand{\eminip}{\end{minipage}}
\newcommand{\btab}{\begin{tabular}}
\newcommand{\etab}{\end{tabular}}

\newcommand{\nnb}{\nonumber}

\newcommand{\hiroshima}{Graduate School of Science, Hiroshima University, Kagamiyama, Higashi-Hiroshima 739-8526, Japan}

\newcommand{\izest}{International Center for Zetta-Exawatt Science and Technology, Ecole Polytechnique, Route de Saclay, Palaiseau, F-91128, France}
\newcommand{\icr}{Institute for Chemical Research, Kyoto University Uji, Kyoto 611-0011, Japan}
\newcommand{\inrs}{Institut national de la recherche 
scientifique-Energie Materiaux Telecommunications 1650 
boulevard Lionel-Boulet,C.P.1020, Varennes (Quebec) J3X 1S2, Canada}
\newcommand{\kyoto}{Graduate School of Science, Kyoto University, Sakyouku, Kyoto 606-8502, Japan}

\begin{document}
\input{body_20150408_mod.tex}

\input{ref_20150408.tex}
\end{document}

%% file: body_20150408_mod.tex
\title{Search for sub-eV scalar and pseudoscalar resonances via four-wave mixing with a laser collider}

\author{Takashi Hasebe} \affiliation{\hiroshima}
\author{Kensuke Homma\footnote{homma@hepl.hiroshima-u.ac.jp}} \affiliation{\hiroshima}\affiliation{\izest}
\author{Yoshihide Nakamiya} \affiliation{\icr}
\author{Kayo Matsuura} \affiliation{\hiroshima}
\author{Kazuto Otani} \affiliation{\inrs}
\author{Masaki Hashida} \affiliation{\icr}\affiliation{\kyoto}
\author{Shunsuke Inoue} \affiliation{\icr}\affiliation{\kyoto}
\author{Shuji Sakabe} \affiliation{\icr}\affiliation{\kyoto}

\date{\today}

\begin{abstract}
\parindent = 0pt
\par The quasi-parallel photon-photon scattering by combining two-color laser fields is an approach to produce resonant states of low-mass fields in laboratory. In this system resonances can be probed via the four-wave mixing process in the vacuum. A search for scalar and pseudoscalar fields was performed by combining 9.3  $\mu$J/0.9 ps Ti-Sapphire laser and 100 $\mu$J/9 ns Nd:YAG laser. No significant signal of four-wave mixing was observed.  We provide the upper limits on the coupling-mass relation for scalar and pseudoscalar fields, respectively, at a 95\% confidence level in the mass region below 0.15~eV. 
\end{abstract}

\maketitle

\section{Introduction}
Uncovering the nature of dark energy and dark matter is one of the most crucial problems in modern physics. 
Low-mass and weakly coupling fields predicted by theoretical models in cosmology and particle physics can be candidates for such dark components.   
For instance, based on the scalar-tensor theory with the cosmological constant $\Lambda$ (STT$\Lambda$)~\cite{STTL1}, 
dark energy is interpreted as decaying $\Lambda$ while the universe becomes older
due to the gravitational coupling between extremely light dilatons, 
a kind of scalar fields ($\phi$), and matter fields. 
Observing the $\gamma\gamma\to\phi\to\gamma\gamma$ process with extremely high intensity laser fields can be a method of searching for $\phi$ in laboratory~\cite{DEptp}. The same approach can also be applied to searches for low-mass pseudoscalar fields ($\sigma$), if the photon spin states are properly chosen~\cite{DEptep}.  Axion \cite{Axion1,Axion2}, a pseudoscalar field associated with breaking of Peccei-Quinn symmetry \cite{PQ}, is a suitable candidate to which this method is directly applicable. Axion is supposed to be one of the most reasonable candidates for cold dark matter~\cite{AxionDM1,AxionDM2}. Therefore, these theoretical models strongly motivate us to search for such fields in laboratory in general. 

Axion searches via the two photon coupling processes have been performed 
by a number of experiments, for example, solar axion searches \cite{sumico1,sumico2,sumico3,cast_vacuum1,cast_vacuum2,cast4he,cast3he}, light shining through a wall \cite{BRFT,PVLAS,OSQAR,alps},  and the axion dark matter experiment \cite{admx1,admx2}.     
Following the first search for scalar fields at quasi-parallel colliding system (QPS)~\cite{hiroshima},
the upgraded search for sub-eV scalar and pseudoscalar fields is presented in this paper.

%
\begin{figure}
\begin{center}
\includegraphics[scale=0.4]{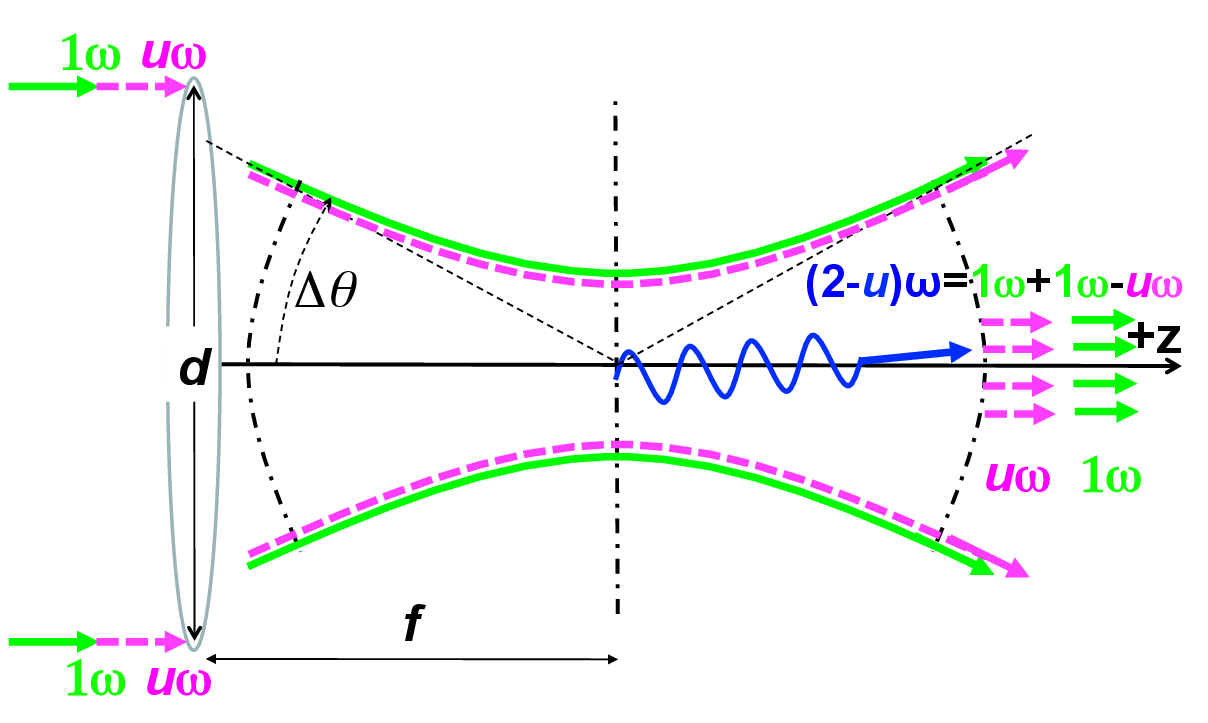}
\end{center}
\caption{Quasi-parallel colliding system by combining two-color laser fields~\cite{DEptp}, where beam diameter $d$, focal length $f$, and the incident angle $\vartheta$ takes $0<\vartheta\leq \Delta\theta$ which is unavoidable due to ambiguity of the wave vectors of incident photons by the nature of focused lasers. }
\label{qps}
\end{figure}

With the schematic view of QPS in Fig. \ref{qps}, we briefly explain the essence
of our method as follows. By using variables defined at QPS,
the center of mass system (CMS) energy between a randomly selected photon pair is expressed as  
\beq
E_{CMS}=2\omega\sin{\vartheta},
\eeq
where $\omega$ is the energy of incident photons and $\vartheta$ is half of the incident angle of the photon pair. Extremely low collision energies are realizable at QPS by focusing a laser field because small values of $\vartheta$ can be automatically introduced. 

In order to overcome low scattering amplitudes of
$\gamma\gamma\to\phi/\sigma\to\gamma\gamma$ processes due to weak coupling, 
we first utilize the character of the integrated resonance effect by capturing
$E_{CMS}$ within $\Delta E_{CMS}$ via $\Delta\theta$ prepared by a creation laser field.
Secondly we let another laser field propagate into the optical axis common 
to the creation laser. 
This laser induces decay of resonance states into a specific energy-momentum space
by the coherent nature of the inducing field. The scattering probability is thus
proportionally increased by the number of photons in the inducing laser field
~\cite{DEptp,TajimaHomma,DEapb,DEptep}.

Energies of decayed photons are defined by the following energy conservation 
\beq
\omega + \omega = (2-u)\omega + u\omega,
\label{fwm}
\eeq
where $u$ is an arbitrary number which satisfies $0<u<1$.  We re-define the energies of final state photons as following   

\beqa 
\omega_{3} &\equiv& (2-u)\omega, \nonumber \\
\omega_{4} &\equiv& u\omega,  
\eeqa
where $\omega_3$ and $\omega_4$ are energies of 
signal photon and inducing photons, respectively.  

In the case of the scalar field exchange, the relation of linear polarization states between initial and final state photons when the wave vectors are on the same reaction plane are expressed as follows:
\beqa
\label{pol_scalar}
\omega\{1\} + \omega\{1\} = &\omega_{3}\{1\} + \omega_{4}\{1\},  \nonumber \\
\omega\{1\} + \omega\{1\} = &\omega_{3}\{2\} + \omega_{4}\{2\}, 
\eeqa
\par\parindent = 0pt
where \{1\} and \{2\} are linear polarization states orthogonal to each other. 
In the pseudoscalar filed exchange, the polarization relation are expressed as    

\beqa
\label{pol_pseudoscalar}
\omega\{1\} + \omega\{2\} = &\omega_{3}\{1\} + \omega_{4}\{2\}, \nonumber \\
\omega\{1\} + \omega\{2\} = &\omega_{3}\{2\} + \omega_{4}\{1\}.
\eeqa
We emphasize that above relations are limited only to the theoretically ideal case 
where all four photons are on the same reaction plane within the treatment based on plane waves.
In the focused QPS, however, 
we must accept independent rotations of the incident $p_1-p_2$ plane and the outgoing $p_3- p_4$ plane
as illustrated in Fig. \ref{Fig9} with respect to an experimentally given linear polarization plane.
This implies that even if we supply $\omega$ as the pure \{1\}-state by a polarizer 
at the moment of plane wave propagation in advance of focusing, 
mixing of \{1\} and \{2\} states for randomly selected incident photon pairs
is unavoidable while lasers are focused. Therefore, the focused QPS with a fixed initial linear
polarization plane has sensitivity to both scalar and pseudoscalar fields simultaneously.
We discuss about this nature in detail in Appendix A.

\parindent = 5pt
The relation in Eq.(\ref{fwm}) is similar to "four-wave mixing" in matter corresponding to the third order non-linear quantum optical process in atoms~\cite{FWM, Yariv}. Therefore, the observation of the four-wave mixing process in the vacuum may be interpreted as a replacement of the atomic nonlinear process by the exchange of unknown scalar or pseudoscalar fields. The observation of four-wave mixing in the vacuum is also used as a method for testing higher-order QED effect~\cite{FWMqed1,FWMqed2,FWMqed3,FWMqed4}.  

\parindent = 5pt
Photons produced via the atomic four-wave mixing process 
can be the main background source for this search.  
The first search for scalar fields at QPS~\cite{hiroshima} was performed with weak intensity lasers, thus, the effect of the four-wave mixing process in atoms was negligible.  
In this experiment, however, the four-wave mixing photons originating from the residual gas are
anticipated due to much higher beam intensities. 
In this paper the method to obtain the exclusion limits in 
the search at QPS sensitive to both scalar and pseudoscalar fields is provided
under the circumstance where a finite amount of background photons must be evaluated. 

\section{The coupling-mass relation}
The effective interaction Lagrangians coupling between two photons and $\phi$ / $\sigma$ are expressed as
\beq
-L_{\phi} = gM^{-1}\frac{1}{4}F_{\mu\nu}F^{\mu\nu}\phi , \hspace{10pt} -L_{\sigma} = gM^{-1}\frac{1}{4}F_{\mu\nu}\tilde{F}^{\mu\nu}\sigma,
\label{eq_phisigma}  
\eeq
where $M$ has the dimension of energy and $g$ is the dimensionless constant. The yield of signal photons, $\mathcal{Y}$, is expressed with experimental parameters relevant 
to lasers and optical elements as follows:   

\beq
\mathcal{Y}= \frac{1}{64\sqrt{2}\pi^{4}}   \left(\frac{\lambda_{c}}{c\tau_{c}}\right) \left(\frac{\tau_c}{\tau_i}\right)    \left(\frac{f}{d}\right)^{3}   \tan^{-1} \left(\frac{\pi d^{2}}{4f\lambda_{c}}\right)      \frac{(\overline{u}-\underline{u})^2}{\overline{u}\underline{u}}  \left(\frac{gm[\rm{eV}]}{M[\rm{eV}]}\right)^2    \left(\frac{m[\rm{eV}]}{\omega[\rm{eV}]}\right)^3 \mathcal{W}  \mathcal{G} \mathcal{F}_{s} C_{mb}  {N_{c}}^2   N_{i},
\eeq
where the subscripts $c$ and $i$ indicate creation and inducing laser, respectively, $\lambda$ is wavelength, $\tau$ is pulse duration, $f$ is focal length, $d$ is beam diameter, $\overline{u}$ and $\underline{u}$ are upper and lower values on $u$ determined by the spectrum width of $\omega_{4}$, respectively, $m$ is mass of the exchanging field, $\mathcal{W}$ is the numerical factor relevant to the integral of the weighted resonance function which is refined in Eq.(\ref{eq_calWnew}) in Appendix B compared to $\mathcal{W} \sim \pi/2$ in Ref.\cite{hiroshima}, $\mathcal{G}$ is the incident-plane-rotation factor described in Appendix A, $\mathcal{F_{S}}$ is the polarization dependent axially asymmetric factor for outgoing photons~\cite{DEptep}, $C_{mb}$ is the combinatorial factor originating from selecting a pair of photons among multimode frequency states and $N$ is the average numbers of photons in the coherent state.
The detail of the formulation of the signal yield is summarized in Appendix of Ref.\cite{hiroshima}. 
The coupling constant $g/M$ is expressed as

\beq
\frac{g}{M[\mbox{eV}]}= 2^{1/4} 8\pi^{2} \sqrt{ \frac{ \mathcal{Y} \omega^3[\rm{eV}] }  { \left(\frac{\lambda_{c}}{c\tau_{c}}\right) \left(\frac{\tau_c}{\tau_i}\right)  \left(\frac{f}{d}\right)^{3}    \tan^{-1} \left(\frac{\pi d^{2}}{4f\lambda_{c}}\right)    \frac{(\overline{u}-\underline{u})^2}{\overline{u}\underline{u}}   \mathcal{W}  \mathcal{G} \mathcal{F}_{s}  C_{mb}  {N_{c}}^2   N_{i} }  }  m^{-5/2} [\rm{eV}].
\label{coupling}
\eeq

\section{Experimental setup}
\begin{figure}[!h]
\begin{center}
\includegraphics[scale=0.3]{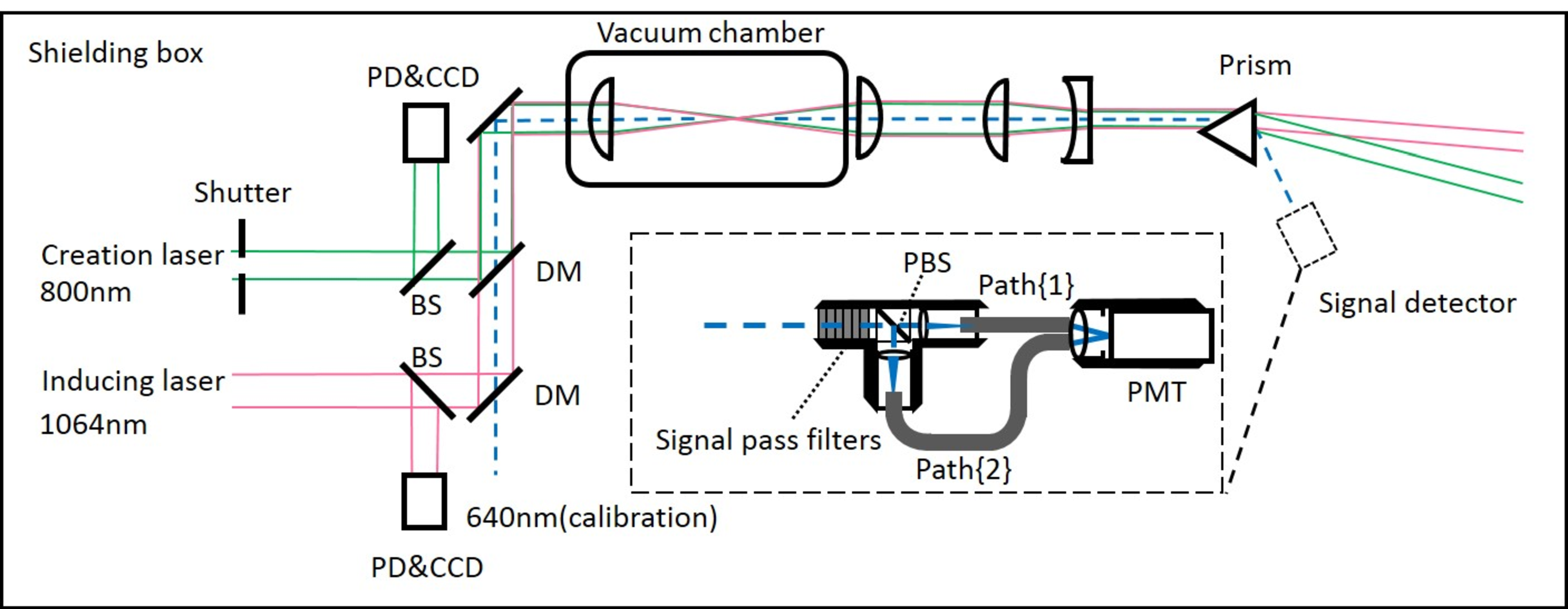}
\end{center}
\caption{The schematic view of the experimental setup
}\label{setup1}
\end{figure}

We explain the experimental setup to detect signals of four-wave mixing in the vacuum. The schematic view of the setup is shown in Fig. \ref{setup1}. 
\parindent = 5pt
\par 
A Ti-Sapphire laser (wavelength 800 nm) and a Nd:YAG laser (wavelength 1064 nm) are used as the creation and the inducing lasers, respectively. To reduce the number of background photons emitted from the residual gas via four-wave mixing, the linear polarization states of the creation and inducing lasers are configured to linear polarization states \{1\} and \{2\}, respectively. The beam alignments of the lasers are monitored by CCD cameras (CCD) and the pulse energies of the creation and inducing lasers are measured by photo-diodes (PD).  These beams are  combined by a dichroic mirror (DM). The combined beams are guided into the vacuum chamber at the 20 mm beam diameter and focused with the convex lens at the focal length 200 mm.

\par The expected wavelength of the corresponding signal photon is evaluated from the following equation 
\beq 
\lambda_s=\frac{{\lambda}_i {\lambda}_c/2} {{\lambda}_i-{\lambda}_c/2} = 641\hspace{3pt} \mbox{nm}.
\eeq
  
\parindent = 0pt
A light source with the central wavelength of 640 nm is combined with the creation and inducing lasers by DM to evaluate the detection efficiency  and  to trace the trajectory of signal photons for the detector alignment.
\parindent = 5pt
\par 
 
\begin{figure}[!h]
  \begin{center}
    \begin{tabular}{c}
      
      \begin{minipage}{0.5\hsize}
        \begin{center}
          \includegraphics[scale=0.42]{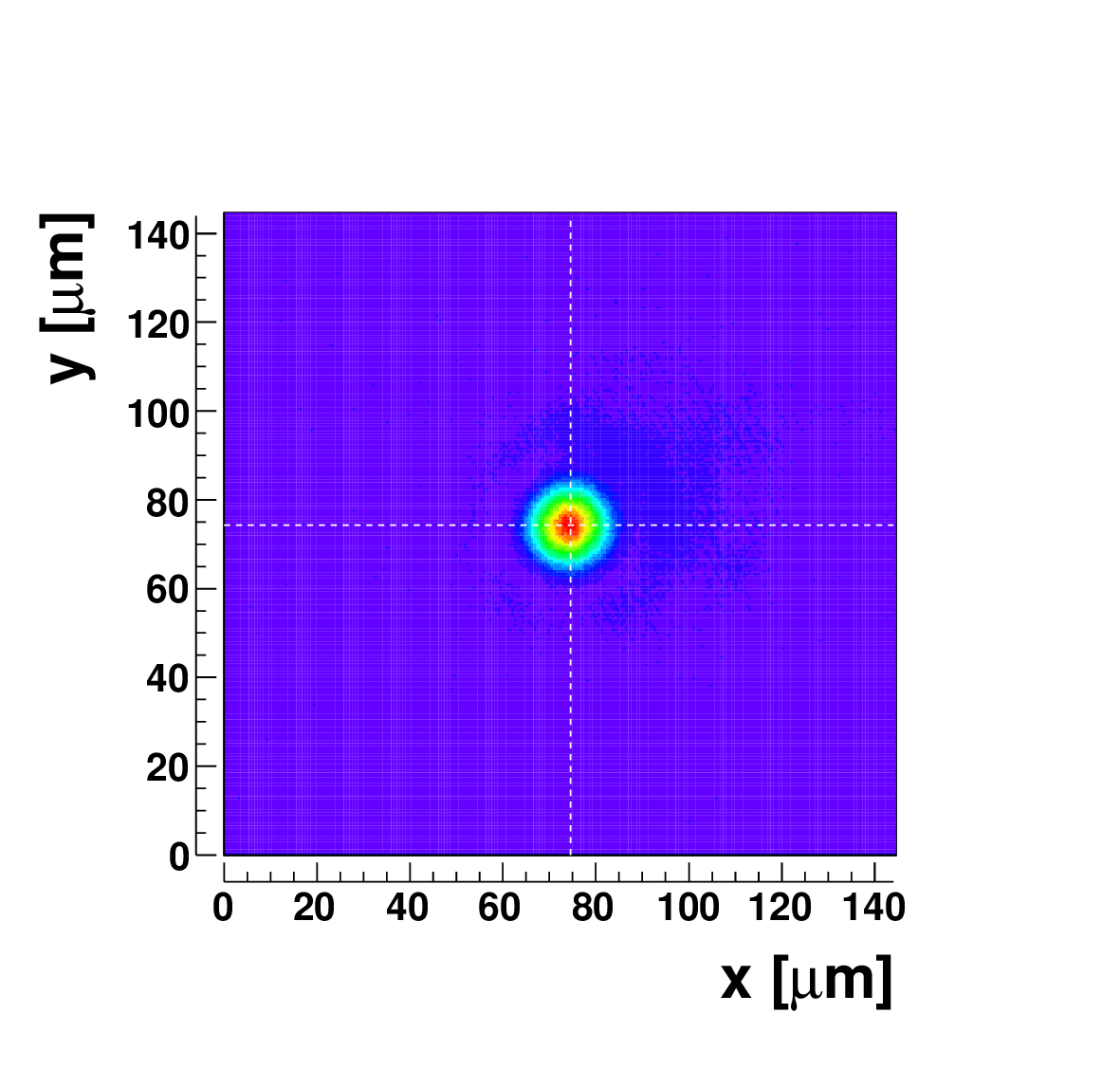}
        \end{center}
      \end{minipage}
    
      \begin{minipage}{0.5\hsize}
        \begin{center}
          \includegraphics[scale=0.42]{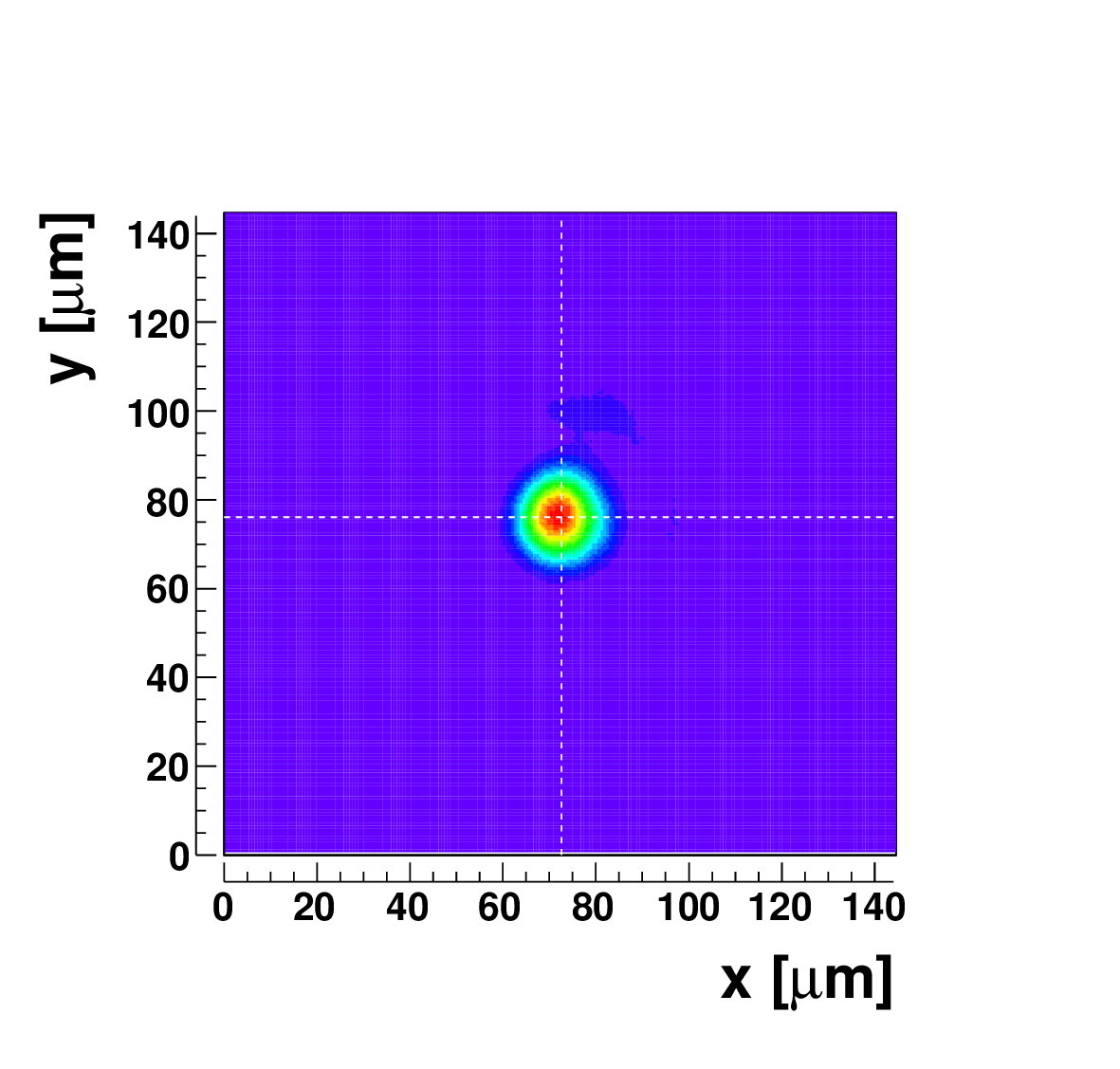}
        \end{center}
      \end{minipage}
    \end{tabular}
\caption{The beam profiles of the creation laser (left) and the inducing laser (right) at the common focal point captured by a common CCD camera.}
\label{spot}
    
  \end{center}
\end{figure}

The agreement of the optical axes between the two lasers are adjusted  at a precision of 2-3 $\mu$m by monitoring individual beam profiles at the near side and the far side of the focal spot with the CCD camera. The beam profiles at the focal spot are shown in Fig. \ref{spot}. The spot sizes of the creation and inducing lasers which are defined as 2 $\sigma$ of the 2D Gauss functions fitting the beam profiles, are 21 $\mu$m and 23 $\mu$m, respectively.  The creation laser overlaps with 87 \% of the beam energy of the inducing laser at the focal spot. Thus, the effective beam energy of the inducing laser is evaluated by correcting the measured beam energy with this overlapping factor.     
 
\par 

\par Signal photons generated within the focal volume travel along the common optical axis of the combined lasers. Signal photons are separated from the creation and inducing lasers by the prism and signal wave filters are placed to further eliminate the residual photons from the combined lasers.  The polarization beam splitter (PBS) transmits \{1\}-polarized photons and reflects \{2\}-polarized photons. Incident photons are split between the shorter optical fiber Path\{1\} and the longer Path\{2\}. The incident photons to PBS are eventually observed by the common photo-device having relative time delay of 23 ns. We use a single-photon-countable photomultiplier tube (PMT) R7400-01 manufactured by HAMAMATSU as the photo-device.

\par 

The repetition rate of the creation laser is 1 kHz and that of the inducing laser is 10 Hz by synchronizing the trigger with the 1 kHz creation pulsing. The data acquisition trigger of 20 Hz is synchronized with the 1 kHz creation laser pulsing which includes pedestal triggers in order to provide four patterns of triggers. The time coincidence between creation and inducing pulses are performed by adjusting the relative injection timing between  the two lasers so that the relative time maximizes the four-wave mixing yield in the air. 
The shutter is placed on the creation laser beam line and it repeats open and close every 5 sec. We acquire data with the four patterns of triggers, which are "both of lasers are incident (S)", "only creation laser is incident (C)", "only inducing laser is incident (I)", and "neither of lasers are incident (P)". The digital oscilloscope recorded waveform data from the PMT and two photo-diodes synchronized with the 20 Hz data acquisition trigger. The recorded waveform data from the PMT are sorted into four types of trigger patterns S, C, I and P. The four trigger patterns are classified by checking the charge correlations between the waveform data from the two photo-diodes for intensity monitoring.

\section{Method of the waveform analysis}

\parindent = 0pt

\begin{figure}[!h]
\begin{center}
\includegraphics[scale=0.6]{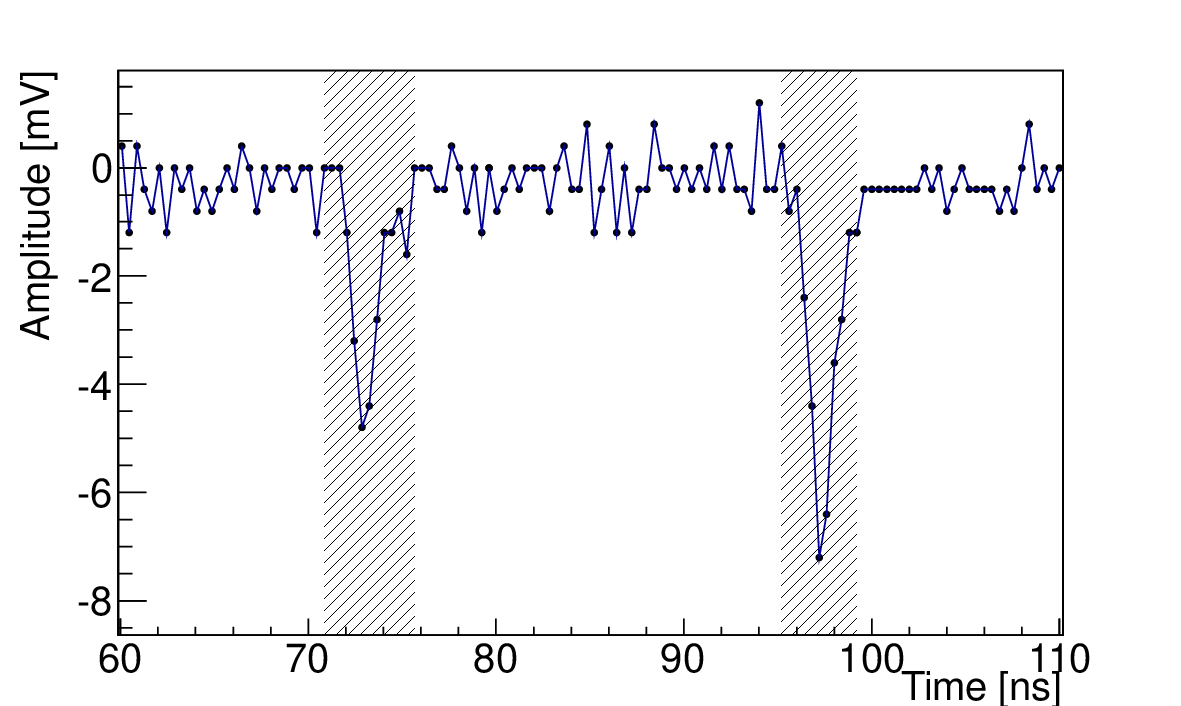}
\end{center}
\caption{ The waveform data sample which has two peak structures. Black shaded areas show the integral ranges to evaluate charge sums of individual peak structures.}
\label{wfm_sample}
\end{figure}

The observed photon counts are estimated by analyzing the waveform data from the PMT. The individual waveform consists of 500 sampling data points within a 200 ns time window. We search for negative peaks of which amplitude exceed a given threshold. We then calculate charge sums of the peak structures. Figure \ref{wfm_sample} shows a sample of waveform data where peak structures are identified. Charge sums of peak structures are evaluated in units of the single-photon equivalent charge, $-4.21\times 10^{-14}$C.     

\parindent = 5pt
\par
\begin{figure}[!h]
\begin{center}
\includegraphics[scale=0.8]{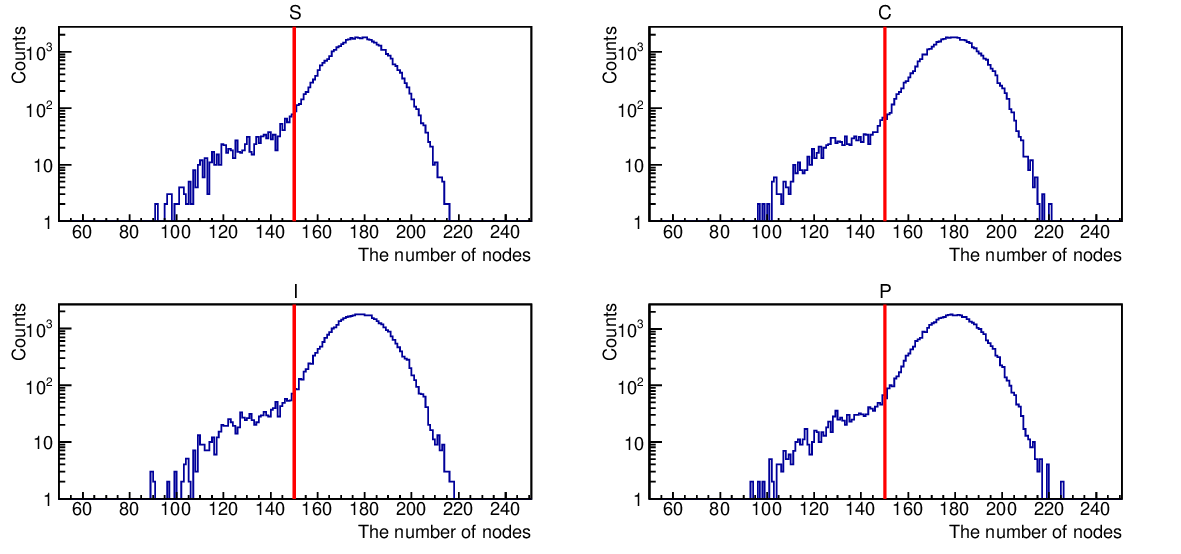}
\end{center}
\caption{Distributions of the number of nodes for trigger patterns S,C,I and P. The events with the fewer number of nodes below the red vertical line are identified as noisy events.}
\label{nodes}
\end{figure}

\begin{figure}[!h]
\begin{center}
\includegraphics[scale=0.8]{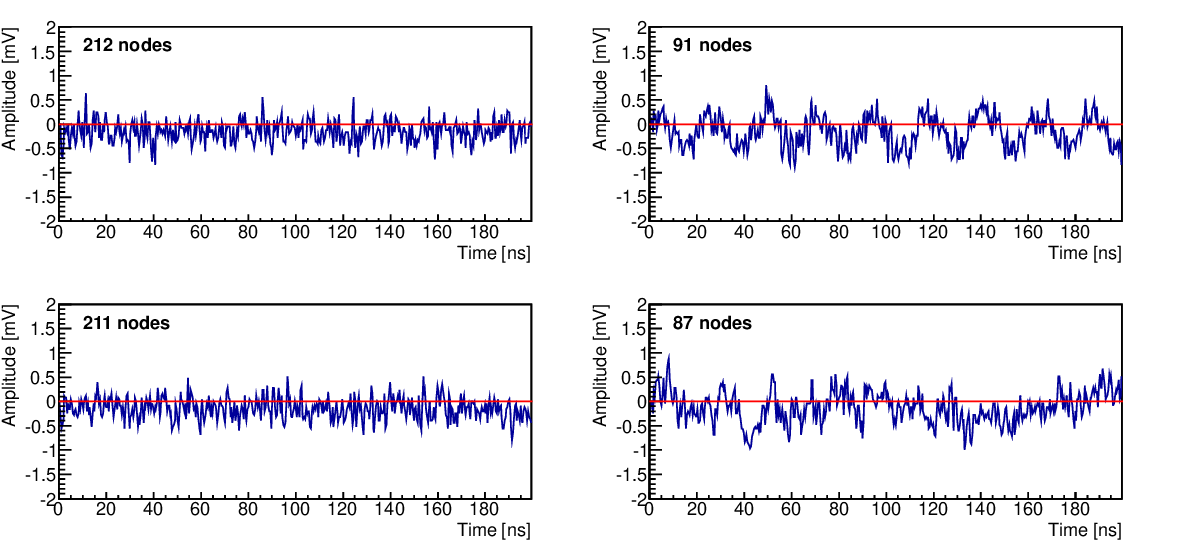}
\end{center}
\caption{ Examples of waveforms of noisy events and normal events. The two panels in the left and right sides show waveforms of normal events and noisy events, respectively. The red horizontal lines indicate the averages of the amplitudes of sampling points for each waveform.}
\label{noise_sample}
\end{figure}

 There are some accidental noisy events among recorded waveform data.  In our analysis method, these noise structures could be misidentified as large photon-like peak structures. Therefore, it is necessary to remove such noisy events from analyzed waveform data before counting photon-like peaks. We can identify noisy events by analyzing the frequencies of waveforms.  Noisy waveforms tend to have lower frequencies than those of normal waveforms.  The frequencies are estimated by counting the number of nodes which is defined as the intersections between a waveform and the average line of amplitudes within the 200 ns time window. The distributions of the number of nodes for each trigger pattern are shown in Fig. \ref{nodes}. We regard a waveform of which the number of nodes is lower than 150 as a noisy event in all trigger patterns by confirming that the differences of the distributions among four trigger patterns are not prominent. 
The typical waveforms of noisy events and normal events identified by this method are shown in Fig. \ref{noise_sample}.

\section{ Measurement of the four-wave mixing process in the residual gas}
\parindent = 0pt

\begin{figure}[!h]
\begin{center}
\includegraphics[scale=0.8]{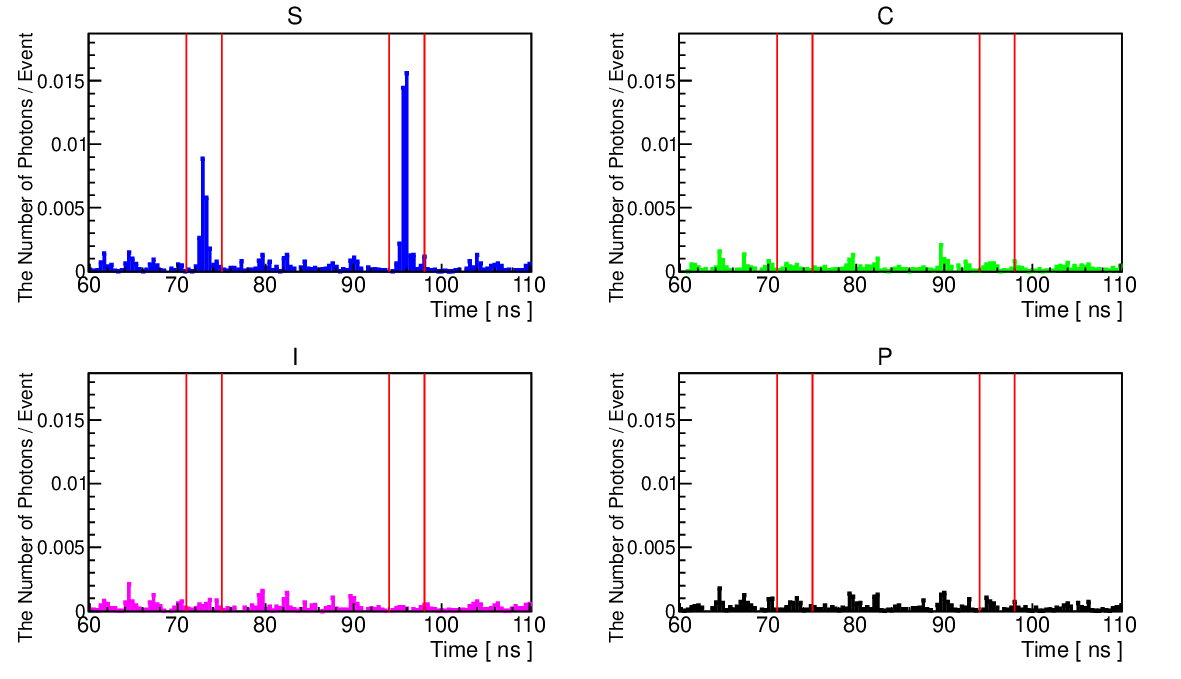}
\end{center}
\caption{The arrival time distributions of observed photons per triggered event (efficiency-uncorrected) at $5.0\times 10^{4}$ Pa. The left and right bands bounded by neighboring two red lines in each panel indicate the time domains $T\{1\}$ and $T\{2\}$ where \{1\} and \{2\}-polarized photons are expected to be observed, respectively. In this figure, the threshold value for peak identification are set lower than that of the actual data analysis on purpose to show typical pedestal structures in each trigger pattern.}
\label{peak_atm}
\end{figure}

The background photons can be produced via the four-wave mixing process occurred  in residual atoms in the vacuum chamber.
To estimate the expected number of background photons, we measured the pressure dependence of the number of four-wave mixing photons in gas. Figure \ref{peak_atm} shows arrival time distributions of observed photons in the air at $5.0\times 10^{4}$ Pa among four trigger patterns. Specific two peak structures appear only at S-pattern. These peak structures have approximately 23ns time interval, which agrees with the optical path-length difference between Path\{1\} and Path\{2\}. We count the number of photons within a time domain $T\{1\}$ (71-75 ns) with \{1\}-polarized state and  $T\{2\}$   (94-98 ns) with \{2\}-polarized state. 

The number of four-wave mixing signals $N_{S}$ are evaluated from the following equation. ( see Eqs.(18) and (19) in Ref.\cite{hiroshima})
\beq 
N_{S} = n_{S} - \frac{W_{S}}{W_{C}}n_{C} - \frac{W_{S}}{W_{I}}n_{I} + \frac{W_{S}}{W_{P}}n_{P},
\label{sub}	
\eeq
where $n_{i}$ and $W_{i}$ denote the number of photon-like peaks in the signal domains and the number of events in trigger pattern $i$, respectively.   
    
\begin{figure}[!h]
\begin{center}
\includegraphics[scale=0.6]{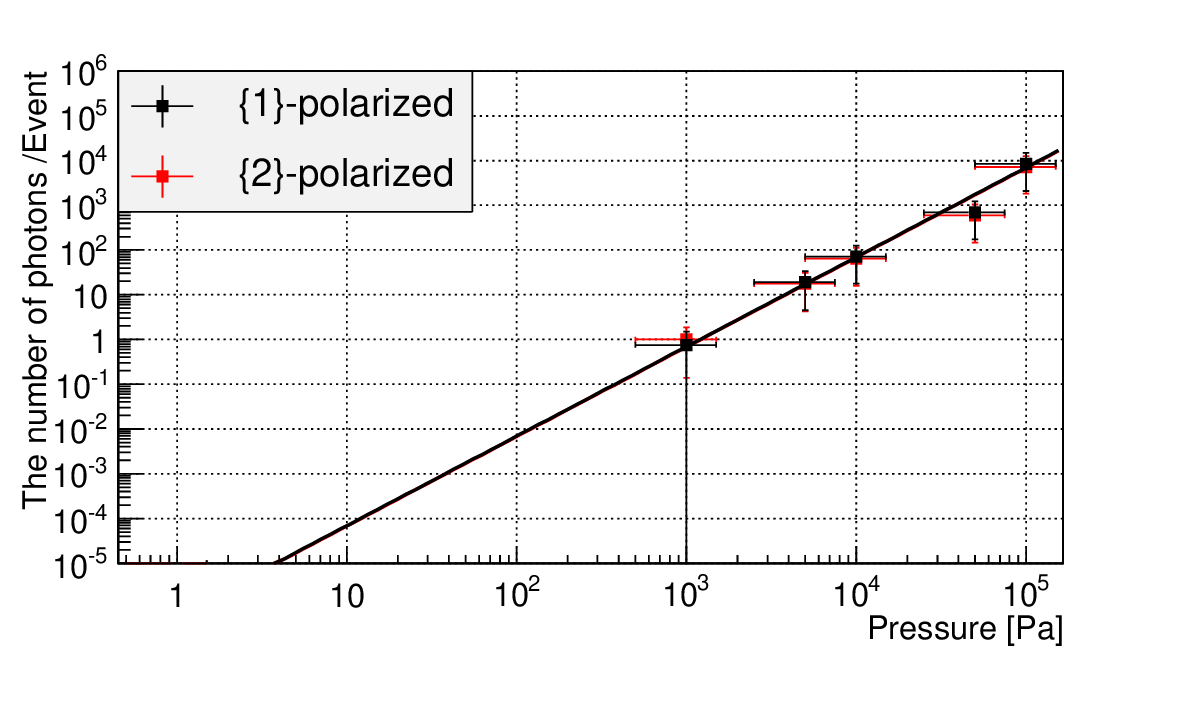}
\end{center}
\caption{The pressure dependence of the number of four-wave mixing photons in the residual gas inside the interaction chamber per S-trigger event. The red and black lines represent the fitting functions for \{1\} and \{2\}-polarized states, respectively.}
\label{pressure}
\end{figure}

The pressure dependence of the number of four-wave mixing photons per S-trigger event are shown in Fig. \ref{pressure}. Data points are fit by the quadratic function of pressure. We extrapolate the number of four-wave mixing photons in the residual gas at $2.3\times10^{-2}$ Pa (an equivalent condition to the vacuum data we discuss later) from the fitting function. The efficiency-corrected number of \{1\}-polarized and \{2\}-polarized photons in residual gas $\mathcal{N}_{gas1}$ and $\mathcal{N}_{gas2}$ with the same shot statistics as the vacuum data are evaluated as follows:    

\beqa
\mathcal{N}_{gas1} &= 1.7 \pm 1.1 \times 10^{-5}, \nonumber \\
\mathcal{N}_{gas2} &= 1.7 \pm 1.1 \times 10^{-5}.
\label{total_n_gas}
\eeqa


We confirmed that the expected value of four-wave mixing photons from the residual gas are negligibly small in the vacuum data for a given total statistics.

\section{Search for four-wave mixing signals in the vacuum}
\begin{figure}[!h]
\begin{center}
\includegraphics[scale=0.8]{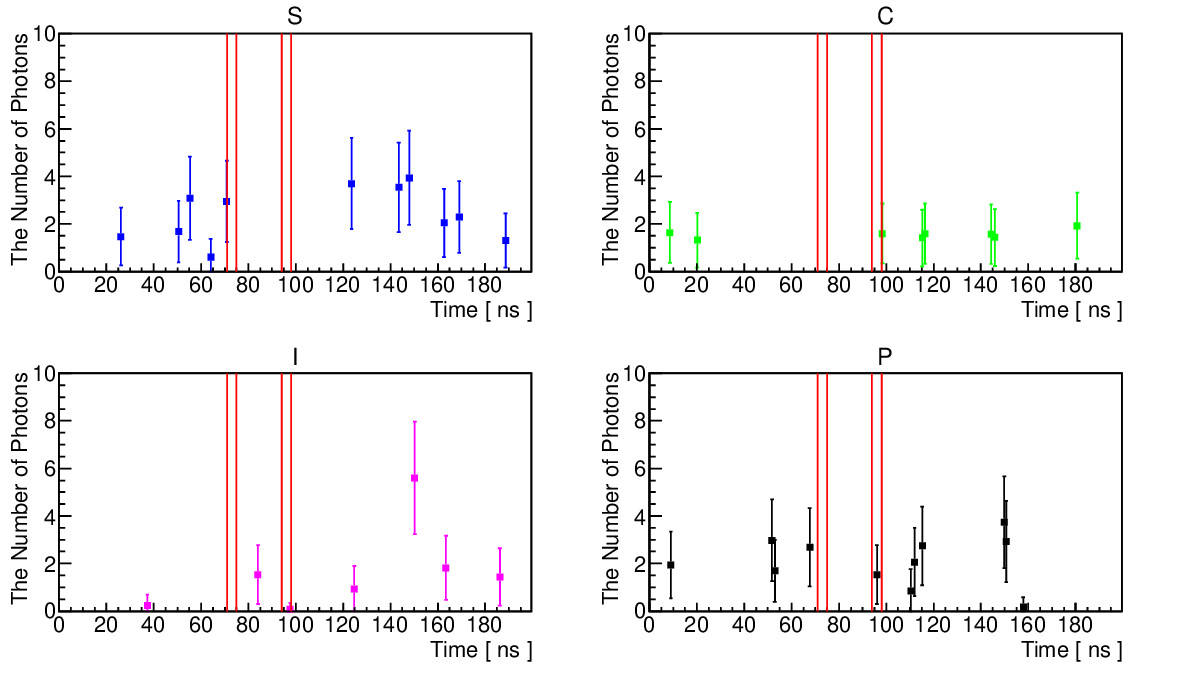}
\end{center}
\caption{Arrival time distributions of observed photons at $2.3\times10^{-2}$ Pa. The data points in each trigger pattern are normalized to the number of triggered events of S-trigger pattern. }
\label{peak_vacuum}
\end{figure}

We acquired data at $2.3\times10^{-2}$ Pa for the search for the resonant states of $\phi$ and $\sigma$ fields. Figure \ref{peak_vacuum} shows the arrival time distributions of observed photon counts . Table \ref{Tab1} summarizes the numbers of observed photon-like signals evaluated in units of the single-photon equivalent charge with \{1\} and \{2\}-polarized states for each trigger pattern, respectively.   

\begin{table}[ht!]
\caption{The numbers of observed photons in $T\{1\}$ and $T\{2\}$ for each trigger pattern. $n_{i1}$  and $n_{i2}$ are the number of photons evaluated in units of single-photon equivalent charge in trigger pattern $i$ with \{1\} and \{2\}-polarized states, respectively.  $W_{i}$ is the number of events in trigger pattern $i$.}
\begin{center}
\begin{tabular}{lccr}  \\ \hline 
 Trigger $i$   & $n_{i1}$  & $n_{i2}$ & $W_{i}$\\  \hline \hline
S & 0 & 0 & 46120\\
C & 0 & 0 & 46203\\ 
I  & 0 & 0.07 & 46044\\  
P & 0 & 1.53 & 46169\\ \hline 

\end{tabular}
\end{center}
\label{Tab1}
\end{table}

\begin{figure}[!h]
\begin{center}
\includegraphics[scale=0.6]{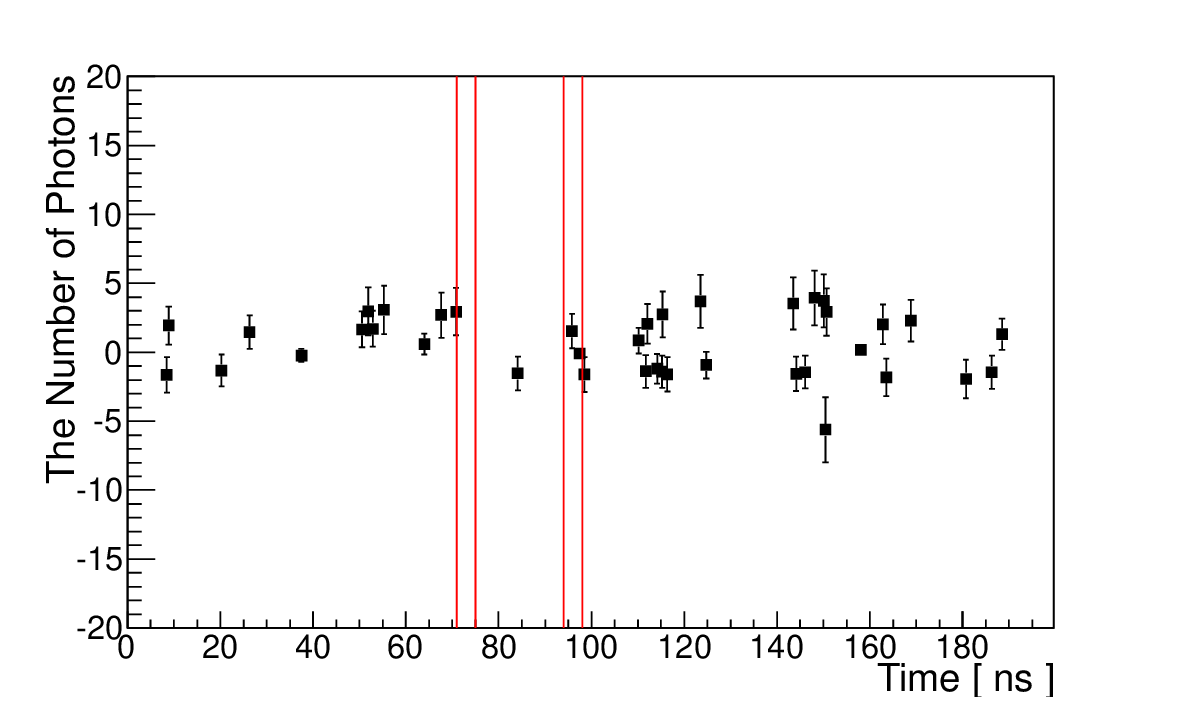}
\end{center}
\caption{The arrival time distribution of $N_{S}$ defined in Eq.(\ref{sub}).}
\label{h_sub}
\end{figure}

After performing subtractions between four patterns of  histograms in Fig. \ref{peak_vacuum} based on the relation in Eq.(\ref{sub}), we obtained the time distribution of $N_{S}$ as shown in Fig. \ref{h_sub}.  
The number of signals with \{1\} and \{2\}-polarized states are, respectively, given as follows:

\beqa
N_{S1} =&&0 \pm 0(\rm{stat.}) \pm 2.16(\rm{syst.I}) \pm 0.30(\rm{syst.I\hspace{-1pt}I}) \pm 0(\rm{syst.I\hspace{-1pt}I\hspace{-1pt}I}),\nonumber \\
N_{S2} =&&1.46 \pm 1.27(\rm{stat.}) \pm 2.16(\rm{syst.I}) \pm 0.04(\rm{syst.I\hspace{-1pt}I}) \pm 3.59(\rm{syst.I\hspace{-1pt}I\hspace{-1pt}I}).
\label{result}
\eeqa  

The systematic error I originates from the number of the photons out side of the two arrival time windows for \{1\} and \{2\}-polarized states. This was evaluated by calculating the root mean square of $N_{S}$ except in the $T\{1\}$ and $T\{2\}$ windows. The systematic error I\hspace{-1pt}I  originates from the dependence on the threshold values for the peak finding $-1.3 \pm 0.1$~mV. The systematic error I\hspace{-1pt}I\hspace{-1pt}I is relevant to the ambiguities of the rejection of noisy events $150 \pm 5$~nodes.           
\par

\section{The excluded coupling-mass limits for scalar and pseudoscalar fields}

\begin{table}[h!]
\caption{Data table of experimental parameters.
$\mathcal{G}^{sc}_{11}$ and $\mathcal{G}^{ps}_{12}$ represent the incident-plane-rotation factor
for scalar and pseudoscalar field exchanges, respectively. 
The evaluation of $\mathcal{G}$ is discussed in Appendix A.
$\mathcal{F}^{sc}_{1122}$ and $\mathcal{F}^{ps}_{1212}$ denote 
the axially asymmetric fractor for scalar and pseudoscalar field exchanges, respectively. 
See the detail in Appendix of Ref.\cite{DEptep}.
}
\begin{center}
\begin{tabular}{lr}  \\ \hline
parameters & values \\ \hline
center of wavelength of creation laser $\lambda_c$   & 800 nm\\
relative line width of creation laser ($\delta\omega/<\omega>)$ &  $7.5\times 10^{-3}$\\
center of wavelength of inducing laser $\lambda_i$   & 1064 nm\\
relative line width of inducing laser ($\delta\omega_{4}/<\omega_{4}>)$ &  $1.0\times 10^{-4}$\\
duration time of creation laser pulse per injection $\tau_{c}$ & 900 fs \\
duration time of inducing laser pulse per injection $\tau_{i}$ & 9 ns \\
creation laser energy per $\tau_{c}$ & 9.3 $\pm$ 1.2 $\mu$J \\
inducing laser energy per $\tau_{i} $ & 100 $\pm$ 1 $\mu$J \\
focal length $f$ & 200 mm\\
beam diameter of laser beams $d$ & 20 mm\\
upper mass range  given by $\theta < \Delta\theta$ & 0.15 eV\\
$u=\omega_{4}/\omega$ & 0.75\\
incident-plane-rotation factor $\mathcal{G}$
 &  $\mathcal{G}^{sc}_{11}$=19/32\\
 & $\mathcal{G}^{ps}_{12}$=1/2\\
axially asymmetric factor $\mathcal{F}_{s}$
 & $\mathcal{F}^{sc}_{1122}$=19.4\\
 & $\mathcal{F}^{ps}_{1212}$=19.2\\ 
combinatorial factor in luminosity $C_{mb}$  & 1/2\\ 
single-photon detection efficiency $\epsilon_{D}$ & 1.4 $\pm$ 0.1 \% \\
efficiency of optical path from interaction point to path\{1\} $\epsilon_{opt1}$ & 0.5 $\pm$ 0.1 \% \\
efficiency of optical path from interaction point to path\{2\} $\epsilon_{opt2}$ & 0.9 $\pm$ 0.2 \% \\
$\delta{N}_{s1}$ & 2.2\\
$\delta{N}_{s2}$ & 4.4\\
\hline
\end{tabular}
\end{center}
\label{Tab2}
\end{table}
 
There is no significant four-wave mixing signal in this search from the result in (\ref{result}).
We thus evaluate the exclusion regions on the coupling-mass relation as follows.

We estimate the upper limit on the sensitive mass range as
\beq
m < 2\omega\sin\Delta\theta \sim 2\omega \frac{d}{2f} = 0.15 \mbox{~eV}
\eeq
based on values summarized in Table \ref{Tab2}, where $\vartheta$ in Fig. \ref{qps} varies 
from 0 to $\Delta\theta$ defined by a focal length $f$ and a beam diameter $d$.

The number of efficiency-corrected \{1\}-polarized signal photons $\mathcal{N}_{S1}$  and that of \{2\}-polarized signal photons $\mathcal{N}_{S2}$ are evaluated from the following relations with the experimental parameters

\beq 
\mathcal{N}_{S1} = \frac{N_{S1}}{\epsilon_{opt1}\epsilon_{D}} , \hspace{15pt} \mathcal{N}_{S2} = \frac{N_{S2}}{\epsilon_{opt2}\epsilon_{D}},
\label{yeild}  
\eeq

where $\epsilon_{opt1}$ and $\epsilon_{opt2}$ are the attenuation ratios of the signal photons propagating from the interaction point through Path\{1\} and Path\{2\}, respectively. 

These attenuation factors are composed of the transmittance of optical 
devices and the acceptance of signal paths 
with respect to the actual location of the PMT. 
They are inclusively evaluated by sampling the beam energies of the 640~nm 
calibration light at the focal point and the detection point, respectively, 
and taking the ratio between them.
The matching of beam paths between the calibration light and four-wave
mixing signals are ensured by adjusting the beam center of calibration 
light with respect to those of creation and inducing lasers 
at the near side and the far side of the focal spot, respectively. 
$\epsilon_{D}$ is the signal detection efficiency of the PMT mainly
caused by the quantum efficiency of the device. 
$\epsilon_{D}$ is evaluated using a 532~nm pulse laser 
in advance of the search. 
We evaluate the absolute detection efficiency by splitting 
the 532~nm beam equally and taking the ratio between these energies. 
The one is measured by a calibrated beam energy meter
and the other is measured by that PMT with neutral density filters 
with measured attenuation factors. 
We then corrected the difference of the quantum efficiencies between 532~nm 
and 641~nm lights by taking the relative quantum efficiencies 
provided by HAMAMATSU into account.


\parindent = 5pt
We then evaluate upper limits on the coupling-mass relation at a 95\% confidence level on the basis that the fluctuation of the number of signal yields forms a Gaussian distribution. We define $\delta{N}_{S}$  as the one standard deviation of ${N}_{S}$. It is evaluated from the quadratic sum of statistical and systematic errors in Eq.(\ref{result}) and 2.24$\delta{N}_{S}$ is the upper limit of ${N}_{S}$ when we obtain a 95\% confidence level~\cite{PDGstatistics}.  The upper limit of signal yields per shot $\mathcal{Y}_{sc}$ (for the scalar field exchange) and $\mathcal{Y}_{ps}$ (for the pseudoscalar field exchange) are evaluated as follows:     

\beq
\mathcal{Y}_{sc} = \frac{2.24\delta{N}_{S2}}{\epsilon_{opt2}\epsilon_{D}W_{S}} , \hspace{15pt} \mathcal{Y}_{ps} = \frac{2.24\delta{N}_{S1}}{\epsilon_{opt1}\epsilon_{D}W_{S}}.
\eeq
As we briefly mention in Introduction and in detail in Appendix A, even though we fix linear polarization planes for creation and inducing laser fields by the polarizers at the moment of plane wave propagation, mixing of \{1\} and \{2\}-polarization states is unavoidable in the focused QPS. By this effect, the focused system has sensitivity to both scalar and pseudoscalar fields simultaneously.




\parindent = 5pt
\par
\begin{figure}[!h]
\begin{center}
\includegraphics[scale=0.8]{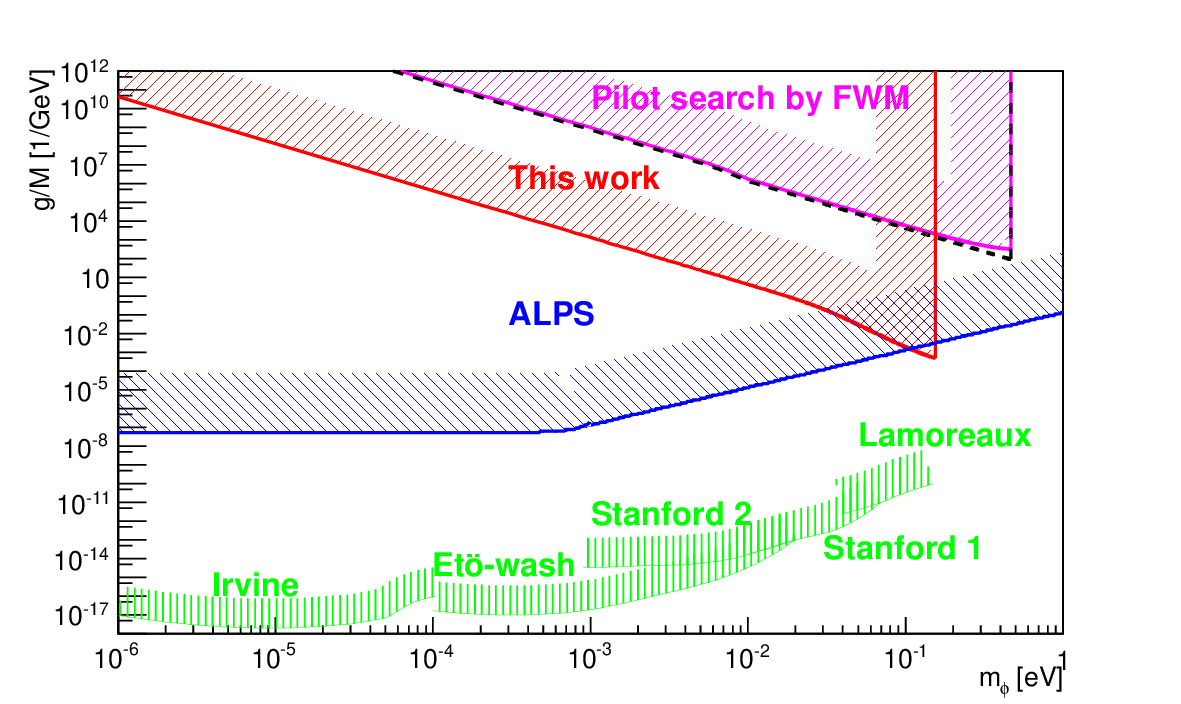}
\end{center}
\caption{Exclusion limits for scalar fields ($\phi$)  in $\phi$-photon coupling ($g/M$) as a function of mass of $\phi$ ($m_{\phi}$). The excluded region by this experiment is drawn by the red shaded area. The magenta shaded area shows the excluded region by our previous search, which is renewed from the black dotted line obtained from Ref.\cite{hiroshima} by taking the incident-plane-rotation factor $\mathcal{G}$ and the mass-dependent $\mathcal{W}$ factor in Appendix B into account. The blue shaded area represents the excluded region for scalar fields by light shining through a wall experiment "ALPS"~\cite{alps} (For the mass region above $10^{-3}$eV,  the sine function part of the sensitivity curve is simplified to unity for drawing purposes). The green shaded areas indicate the limits given by non-Newtonian force searches by torsion balance experiments "Irvine"~\cite{Irvine}, "Eto-wash"~\cite{Eto-wash1,Eto-wash2}, "Stanford1"~\cite{Stanford1}, "Stanford2"~\cite{Stanford2} and Casimir force measurement "Lamoreaux"\cite{Lamoreaux}. } 
\label{coupling_sc}
\end{figure}

\begin{figure}[!h]
\begin{center}
\includegraphics[scale=0.8]{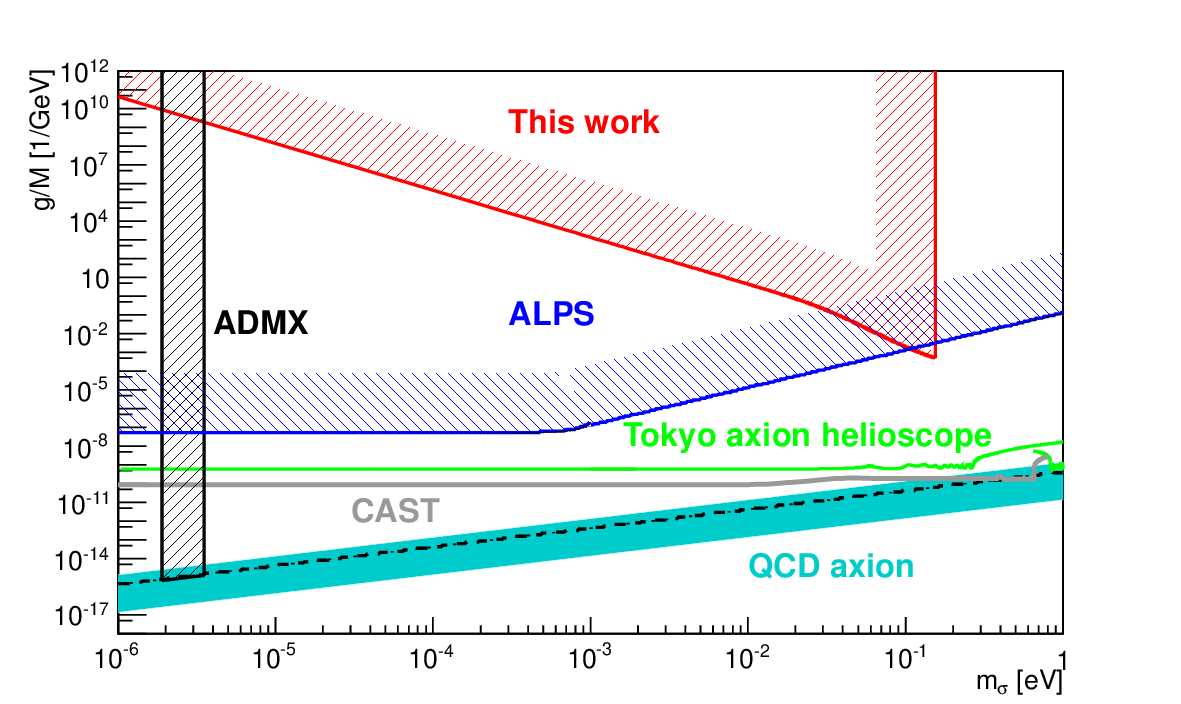}
\end{center}
\caption{Exclusion limits for pseudoscalar fields ($\sigma$)  in $\sigma$-photon coupling ($g/M$) as a function of mass of $\sigma$ ($m_{\sigma}$). The blue shaded area shows the excluded region by the pseudoscalar search, "ALPS".  The green and gray solid line show the exclusion limits from the solar axion experiments "Tokyo Axion Helioscope"~\cite{sumico1,sumico2,sumico3} and "CAST"~\cite{cast_vacuum2,cast4he,cast3he} , respectively.  The black shaded area represents the result from  the dark matter axion search using a microwave cavity "ADMX"~\cite{admx1,admx2}. The cyan band indicates the expected coupling-mass relation of QCD axion predicted by KSVZ model\cite{KSVZ1,KSVZ2} with $|E/N-1.95|$ in the range 0.07-7, furthermore, in the case of $E/N=0$ is shown by the black dotted line. }
\label{coupling_ps}
\end{figure}

 We obtain the coupling-mass relation from Eq.(\ref{coupling}). The exclusion limits for scalar and pseudoscalar fields at a 95\% confidence level are shown in Fig. \ref{coupling_sc} and Fig. \ref{coupling_ps} , respectively. 

\section {Conclusions}
 A search for scalar and pseudoscalar fields via the four-wave mixing precess at QPS  has been performed by focusing 10 $\mu$J/0.9 ps pulse laser and 100 $\mu$J/9 ns pulse lasers.  The number of \{1\} and \{2\}-polarized signal-like photons are $N_{S1} = 0 \pm 0(\rm{stat.}) \pm 2.16(\rm{syst.I}) \pm 0.30(\rm{syst.I\hspace{-1pt} I }) \pm 0(\rm{syst.I\hspace{-1pt}I\hspace{-1pt} I})$ and $N_{S2} =1.46 \pm 1.27(\rm{stat.}) \pm 2.16(\rm{syst.I}) \pm 0.04(\rm{syst.I\hspace{-1pt}I}) \pm 3.59(\rm{syst.I\hspace{-1pt}I\hspace{-1pt}I})$, respectively.  We confirmed that the expected number of four-wave mixing photons in the residual gas are negligibly small by measuring the pressure dependence.  As a result, no significant four-wave mixing signal is observed in this experiment. We obtained the upper limits on the coupling-mass relation for scalar and pseudoscalar fields at a 95$\%$ confidence level, respectively. The most sensitive coupling limits  $g/M = 5.24\times10^{-4} \rm{GeV}^{-1}$ for scalar search and $g/M = 5.42\times10^{-4} \rm{GeV}^{-1}$ for pseudoscalar search are obtained at $m=0.15$ eV. 

 \section*{Acknowledgments}
We gratefully thank S. Tokita and Y. Miyasaka  for operation and maintenance of laser systems.  
We appreciate Y. Inoue providing the data list 
of the sensitive curve for Tokyo axion helioscope.

K. Homma cordially thanks Y. Fujii for the detailed discussions
and careful checks on the spin-dependence of the scattering probability.
He expresses his gratitude to T. Tajima and G. Mourou
for many aspects relevant to this subject.
He also thank for the strong financial supports by the Grant-in-Aid for
Scientific Research no.24654069, 25287060 and 26104709 from MEXT of Japan, 
Collaborative Research Program of Institute for Chemical Research, Kyoto University 
(grant No. 2012-6, No. 2013-56 and No. 2014-72) and the support by MATSUO FOUNDATION.

\section*{Appendix A: Evaluation of the incident-plane-rotation factor ${\mathcal G}$}

\begin{figure}[h!]
\bcent
\includegraphics[width=13.0cm]{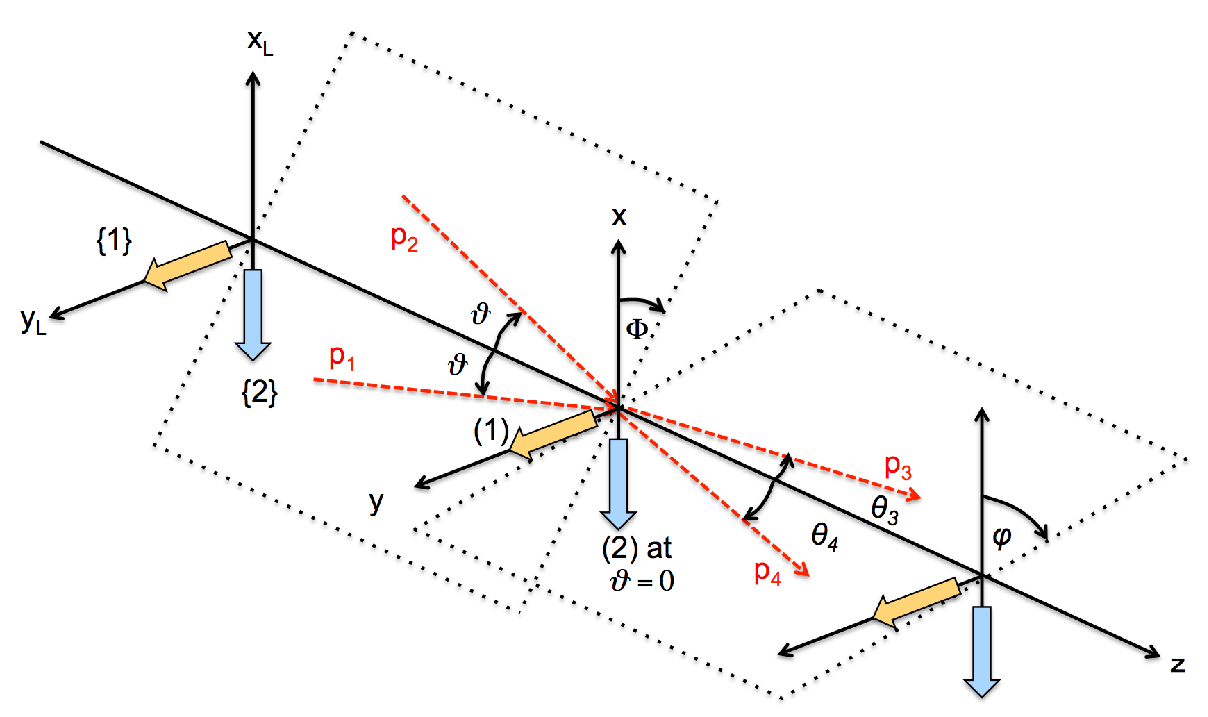}
\caption{
Definitions of polarization vectors and rotation angles in QPS.
}
\label{Fig9}
\ecent
\end{figure}

Figure \ref{Fig9} illustrates the relation between experimentally defined
linear polarization directions \{1\} and \{2\} and 
those theoretically defined (1) and (2). It also depicts relations between $p_1-p_2$
and $p_3-p_4$ planes with respect to the $x-z$ plane 
where the theoretically allowed coupling of an exchanged field 
to the linear polarization states can be evaluated in the clearest way.
In Ref.\cite{DEptep}, we have assumed the incident photons $p_1$ and $p_2$ are 
both plane waves with different wave vectors on the same reaction plane
which always ensures the clearest condition. 
In the general 3-dimensional incident case such as a focused Gaussian beam, however, 
a $p_1-p_2$ plane can rotate with respect to the $x-z$ plane, which results in
a deviation from the theoretically clearest condition.
We, therefore, introduce a weighted averaging factor ${\mathcal G}$ 
over the clockwise rotation angle
$\Phi$ of the incident reaction plane with respect to the $x$-axis as follows.

As we have discussed in ref.\cite{DEptep},
the Lorentz invariant $s$-channel scattering amplitude
for Lagrangian defined in Eq.(\ref{eq_phisigma}) have the following basic form
\beq\label{eqA1}
{\mathcal M}_S = -(gM^{-1})^2
\frac{{\mathcal V}^{[1]}_{ab} {\mathcal V}^{[2]}_{cd}}{(p_1+p_2)^2+m^2},
\eeq
where
$S\equiv abcd$ with $a,b,c,d = 1$ or $2$, respectively,
denotes a sequence of four-photon polarization states
and $m$ is the mass of scalar or pseudoscalar field.
With vectors defined below, the vertex factors for the scalar case (SC)
are expressed as
\beqa
{\mathcal V}_{ab}^{[1]SC}&=&(p_1p_2)(e_1^{(a)}e_2^{(b)})-(p_1e_2^{(a)})(p_2e_1^{(b)}),\qquad\qquad\quad\nnb\\
{\mathcal V}_{cd}^{[2]SC}&=&(p_3p_4)(e_3^{(c)}e_4^{(d)})-(p_3e_4^{(c)})(p_4e_3^{(d)}),
\label{eqAscVV}
\eeqa
and these for the pseudoscalar case (PS) are expressed as
\beqa
{\mathcal V}_{ab}^{[1]PS}&=&-\epsilon^{\mu\nu\rho\sigma}p_{1\mu} p_{2\rho} e_{1\nu}^{(a)}e_{2\sigma}^{(b)},\qquad\quad\nnb\\
V_{cd}^{[2]PS}&=&-\epsilon^{\mu\nu\rho\sigma}{p_3}_{\mu} {p_4}_{\rho} e_{3\nu}^{(c)}e_{4\sigma}^{(d)}.
\label{eqApsVV}
\eeqa

We must first take into account the clockwise rotation angle $\varphi$
of $p_3-p_4$ plane with respect to the given $x-z$ plane
independent of the $p_1-p_2$ plane, because these two planes are not coplanar
in QPS contrary to the situation where the coplanar condition of $p_1$ through $p_4$ is 
always satisfied in CMS. This implies that the simple summation factor $2\pi$
on the azimuthal degree of freedom of solid angle cannot be applied to QPS,
instead, the $\varphi$-dependent squared transition amplitude must be
summed over the possible rotation $\varphi$ from 0 to $2\pi$.
We have already introduced this axially asymmetric factor 
${\mathcal F_S}$ with respect only to
the incident reaction plane at $\Phi = 0$ in \cite{DEptep}.
This factor essentially depends only on the second vertex factors above,
while the incident-plane-rotation factor ${\mathcal G}$ is relevant only to 
the first vertex factors.
We thus define the incident-plane-rotation factor as a weighted average
with respect to ${\mathcal F_S}$ at $\Phi=0$ as follows
\beq\label{eqC00}
{\mathcal G}_{ab} \equiv 
\frac{\int_0^{2\pi} |{\mathcal V}_{ab}^{[1]} (\Phi)|^2 d\Phi}
{\int_0^{2\pi} |{\mathcal V}_{ab}^{[1]} (\Phi = 0)|^2 d\Phi},
\eeq
because experiments cannot fix the incident reaction plane and intensity
of the creation laser field must be shared over possible incident reaction planes.

By requiring (1)=\{1\} and (2)=\{2\} at $\Phi = \varphi =0$ where theoretically
clearest polarization relations can interface with the experimental condition,
we describe the polarization vectors and momentum vectors for four photons
with rotation angles $\Phi$ and $\varphi$ as follows:
\beq\label{eqA4}
e_i^{(1)} = (0, 1, 0),
\eeq
\beqa\label{eqA5}
e_1^{(2)} &=& (-\cos\vartheta, 0, \sin\vartheta), \quad
e_2^{(2)} = (-\cos\vartheta, 0, -\sin\vartheta), 
\nnb\\
e_3^{(2)} &=& (-\cos\theta_3, 0, \sin\theta_3), \quad
e_4^{(2)} = (-\cos\theta_4, 0, -\sin\theta_4), 
\nnb\\
p_1&=&(\omega\sin\vartheta\cos\Phi,-\omega\sin\vartheta\sin\Phi,\omega\cos\vartheta; \omega),\quad\qquad \nnb\\
p_2&=&(-\omega\sin\vartheta\cos\Phi,\omega\sin\vartheta\sin\Phi,\omega\cos\vartheta; \omega),\quad\qquad \nnb\\
p_3&=&(\omega_3\sin\theta_3\cos\varphi,-\omega_3\sin\theta_3\sin\varphi,\omega_3\cos\theta_3; \omega_3), \nnb\\
p_4&=&(-\omega_4\sin\theta_4\cos\varphi,\omega_4\sin\theta_4\sin\varphi,\omega_4\cos\theta_4; \omega_4).
\eeqa
We note here that we cannot rotate polarization vectors because the experiment must
introduce fixed polarization vectors. This implies that the clear distinction
between scalar and pseudoscalar couplings cannot be stated due to
non-zero rotation angles
because non-identical linear polarization planes between photon 1 and 2
or photon 3 and 4 are implicitly introduced.

Based on these vectors,
we summarize relations between momenta and polarization vectors
with photon labels $i=1,2,3,4$ as follows
\beqa\label{eqA6}
\left(p_1 e_j^{(1)} \right) = -\omega\sin\vartheta\sin\Phi, \quad 
\left(p_2 e_j^{(1)} \right) =  \omega\sin\vartheta\sin\Phi, \nnb\\
\left(p_3 e_j^{(1)} \right) = -\omega_3\sin\theta_3\sin\varphi,  \quad
\left(p_4 e_j^{(1)} \right) =  \omega_4\sin\theta_4\sin\varphi,
\eeqa
\beqa\label{eqA7}
\left(e_i^{(1)} e_j^{(1)} \right) = 1 \mbox{\quad and \quad }
\left(e_i^{(1)} e_j^{(2)} \right) = 0
\eeqa
for any pair $i$, $j$, and
\beqa\label{eqA8}
\left(e_i^{(2)} e_j^{(2)} \right)  =  1 \mbox{\quad for\quad } i=j,
\mbox{\qquad \qquad \qquad \qquad \qquad \qquad} \nnb\\
\left(e_1^{(2)} e_2^{(2)} \right)  =  \cos2\vartheta,
\left(e_3^{(2)} e_4^{(2)} \right)  =  \cos(\theta_3+\theta_4)
\equiv\cos\theta_+, \nnb\\
\left(e_1^{(2)} e_3^{(2)} \right)  =  \cos(\vartheta-\theta_3),
\left(e_2^{(2)} e_4^{(2)} \right)  =  \cos(\vartheta-\theta_4), 
\mbox{\qquad} \nnb\\
\left(e_1^{(2)} e_4^{(2)} \right)  =  \cos(\vartheta+\theta_4),
\left(e_2^{(2)} e_3^{(2)} \right)  =  \cos(\vartheta+\theta_3),
\mbox{\qquad}
\eeqa
and
\beq\label{eqA10}
\left(p_1 p_2\right) = \omega^2(\cos2\vartheta-1)
= \left(p_3 p_4\right) = \omega_3\omega_4(\cos\theta_+-1)
\eeq
where $(p_1+p_2)^2 = (p_3 + p_4)^2$ is required for massless photons.

We are now ready to estimate the factor ${\mathcal G}$ included in
the partially integrated cross section at Eq.(A24) in Ref.\cite{hiroshima}.
We evaluate the case of $ab=11$ for the scalar exchange.
From the first of Eq.(\ref{eqAscVV}), we obtain
\beqa\label{eqA11}
{\mathcal V}^{[1]SC}_{11} &=&
(p_1p_2)(e_1^{(1)}e_2^{(1)})-(p_1e_2^{(1)})(p_2e_1^{(1)}) \nnb\\
&=& \omega^2(\cos2\vartheta-1 + \sin^2\vartheta\sin^2\Phi)
\sim \omega^2\vartheta^2(2-\sin^2\Phi),
\eeqa
where
the first of Eq.(\ref{eqA10}), Eq.(\ref{eqA4}), and
$(p_1e_2^{(1)})(p_2e_1^{(1)}) = -(\omega\sin\vartheta\sin\Phi)^2$
are substituted.
The last approximation is based on $\vartheta \sim \vartheta_r\ll 1$.

This yields the following averaging factor on the incident reaction plane
\beq\label{eqC03}
{\mathcal G}_{11}^{SC} =
\frac{\int_0^{2\pi} (2-\sin^2\Phi)^2 d\Phi}{8\pi}
= \frac{19}{32}.
\eeq

We also provide the case of $ab=12$ for the pseudoscalar exchange as follows.
Based on the first of Eq.(\ref{eqApsVV}), 
the first vertex factor with vector definitions above
is expressed as
\beqa
{\mathcal V}_{12}^{[1]PS}
&=&-\epsilon^{\mu\nu\rho\sigma}p_{1\mu}p_{2\rho}e_{1\nu}^{(1)}e_{2\sigma}^{(2)}
=-p_{1\mu}p_{2\rho} \epsilon^{\mu y \rho \sigma}e_{2\sigma}^{(2)} \nnb\\
&=&-p_{1\mu}p_{2\rho}\left[ \rule[-.1em]{0em}{1.2em}
\epsilon^{\mu y\rho x}(-\cos\vartheta) +\epsilon^{\mu y\rho
z}(-\sin\vartheta)\right] \nnb\\
&=& p_{2\rho}\left[ \rule[-.1em]{0em}{1.2em}
\left( p_{10}\epsilon^{0y\rho x} +p_{1z}\epsilon^{zy\rho x}
\right)\cos\vartheta
+\left( p_{10}\epsilon^{0y\rho z} +p_{1x}\epsilon^{xy\rho z}
\right)\sin\vartheta
\right] \nnb\\
&=& p_{2\rho}\left[ \rule[-.1em]{0em}{1.2em}
 \left(-\omega \epsilon^{0y\rho x} + \omega\cos\vartheta \epsilon^{zy\rho x}
 \right)\cos\vartheta  +
\left(-\omega \epsilon^{0y\rho z} + \omega\sin\vartheta\cos\Phi \epsilon^{xy\rho z}
 \right)\sin\vartheta
\right] \nnb\\
&=& \left[ \rule[-.1em]{0em}{1.2em}
\left( -\omega \epsilon^{0yzx}p_{2z} +\omega\cos\vartheta
\epsilon^{zy0x}p_{20} \right) \cos\vartheta +
\left(  -\omega\epsilon^{0yxz}p_{2x}
+\omega\sin\vartheta\cos\Phi\epsilon^{xy0z}p_{20}  \right)\sin\vartheta\right] \nnb\\
&=& \omega^2\left[ \rule[-.1em]{0em}{1.2em}
\left( -\cos\vartheta +\cos\vartheta \right)\cos\vartheta +\left(
-\sin\vartheta  -\sin\vartheta\right)\cos\Phi\sin\vartheta \right]
=-2\omega^2 \sin^2\vartheta\cos\Phi.
\label{eqC05}
\eeqa

This yields the following averaging factor on the incident reaction plane
\beq\label{eqC05}
{\mathcal G}_{12}^{PS} =
\frac{\int_0^{2\pi} \cos^2\varphi d\varphi}{2\pi}
= \frac{1}{2}.
\eeq

\section*{Appendix B: Refinement of the weight factor ${\mathcal W}$}
In Ref.\cite{DEptep,hiroshima}, we approximated ${\mathcal W}$ as 
a constant $\pi/2$ for the mass region much smaller
than that covered by $\Delta\theta$ as a conservative estimate.
This is because we rather respected simplicity of the parametrization
than accuracy.
However, once we need to compare the sensitivity for the higher mass region 
with the other search methods, the validity of the approximation applicable
only to the smaller mass region must be reconsidered.
In the following, we first exactly repeat the relevant part 
of Ref.\cite{hiroshima} and then refine ${\mathcal W}$ as a function
of sensitive mass regions by quoting necessary equations.

We first express the squared scattering amplitude
for the case when a low-mass field is exchanged
in the s-channel via a resonance state with the symbol to describe
polarization combinations of initial and final states $S$.
\beq\label{eq_M2}
|{\mathcal M}_S|^2 \approx  (4\pi)^2 \frac{a^2}{\chi^2+a^2},
\eeq
where
$\chi =\omega^2 -\omega_r^2$
with the resonance condition $m = 2\omega_r\sin\vartheta_r$
for a given mass $m$ and $a$ is expressed as
\beq\label{eq_a}
a = \frac{\omega^2_r}{8\pi}\left(\frac{g m}{M}\right)^2
=\frac{m \Gamma}{2\sin^2\vartheta_r}
\eeq
with the resonance decay rate of the low-mass field
\beq
\Gamma=(16\pi)^{-1} \left( g M^{-1}\right)^2 m^3.
\label{mxelm_4a}
\eeq

The resonance condition is satisfied when the center-of-mass system (CMS)
energy between incident two photons $E_{CMS}=2\omega\sin\vartheta$
coincides with the given mass $m$.
At a focused geometry of an incident laser beam, however,
$E_{CMS}$ cannot be uniquely specified due to the momentum uncertainty
of incident waves. Although the incident laser energy has the intrinsic
uncertainty, the momentum uncertainty
or the angular uncertainty between a pair of incident photons dominates
that of the incident energy.
Therefore, we consider the case where only angles of incidence
$\vartheta$ between randomly chosen pairs of photons
are uncertain within $0<\vartheta\le\Delta\vartheta$
for a given focusing parameter by fixing the incident energy. The treatment
for the intrinsic energy uncertainty is explained in Appendix B later.
We fix the laser energy $\omega$ at the optical wavelength
\beqa\label{exeq_13}
\omega_{opt}^2 = \frac{m^2}{4\vartheta^2_r} \sim 1\mbox{eV}^2,
\eeqa
while the resonance condition depends on the incident angle uncertainty.
This gives the expression for $\chi$ as a function of $\vartheta$
\beqa\label{exeq_14}
\chi(\vartheta) = w^2_{opt} - w^2_r(\vartheta)
= \frac{m^2}{4\vartheta^2_r} - \frac{m^2}{4\vartheta^2}
= \left(1-(\vartheta_r/\vartheta)^2\right) \omega_{opt}^2,
\eeqa
where
\beq\label{eq_dvartheta}
d\vartheta = \frac{\vartheta_r}{2\omega^2_{opt}} (1-\frac{\chi}{\omega^2_{opt}})^{-3/2} d\chi.
\eeq
We thus introduce the averaging process for
the squared amplitude $\overline{|{\mathcal M}_S|^2}$
over the possible uncertainty on incident angles
\beqa\label{eq_M2average}
\overline{|{\mathcal M}_S|^2} = \int^{\pi/2}_{0} \rho(\vartheta)
|{\mathcal M}_S(\vartheta)|^2 d\vartheta
\eeqa
where
${\mathcal M}_S$ specified with a set of physical parameters $m$ and $gM^{-1}$
is expressed as a function of $\vartheta$, and
$\rho(\vartheta)$ is the probability distribution function
as a function of the uncertainty on $\vartheta$ within an incident pulse.

We review the expression for the electric field of the Gaussian laser
propagating along the $z$-direction
in spatial coordinates $(x,y,z)$~\cite{Yariv} as follows:
\beqa\label{eq_Gauss}
\vec{E}(x,y,z) = \vec{E_0}
\qquad \qquad \qquad \qquad \qquad \qquad \qquad \qquad \qquad \nnb\\
\frac{w_0}{w(z)}\exp
\left\{
-i[kz-H(z)] - r^2 \left( \frac{1}{{w(z)}^2}+\frac{ik}{2R(z)} \right)
\right\},
\eeqa
%
where $E_0$ electric field amplitude,
$k=2\pi/\lambda$, $r=\sqrt{x^2+y^2}$, $w_0$ is the minimum waist,
which cannot be smaller than $\lambda$ due to the diffraction limit, and
other definitions are as follows:
%
\beqa\label{eq_wz}
{w(z)}^2 = {w_0}^2
\left(
1+\frac{z^2}{{z_R}^2}
\right),
\eeqa
\beqa\label{eq_Rz}
R = z
\left(
1+\frac{{z_R}^2}{z^2}
\right),
\eeqa
\beqa\label{eq_etaz}
H(z) = \tan^{-1}
\left(
\frac{z}{z_R}
\right),
\eeqa
\beqa\label{eq_zr}
z_R \equiv \frac{\pi{w_0}^2}{\lambda}.
\eeqa

With $\theta$ being an incident angle of a single photon in the Gaussian beam,
the angular distribution $g(\theta)$ can be approximated as
\beq\label{eqGaussAngular}
g(\theta) \sim \frac{1}{\sqrt{2\pi}\Delta\theta}
\exp\left\{-\frac{\theta^2}{2\Delta\theta^2}\right\},
\eeq
where the incident angle uncertainty in the Gaussian beam
$\Delta\theta$ is introduced within the physical range $|\theta|<\pi/2$ as
\beqa\label{eq15}
\Delta\theta
\sim
\frac{\lambda_c}{\pi w_0}
= \frac{d}{2f},
\eeqa
with the wavelength of the creation laser $\lambda_c$, the beam diameter $d$,
the focal length $f$, and the beam waist $w_0 = \frac{f\lambda_c}{\pi d/2}$
as illustrated in Fig.\ref{qps}.
For a pair of photons 1, 2 each of which follows $g(\theta)$, the incident
angle between them is defined as
\beq\label{eqVartheta}
\vartheta = \frac{1}{2}|{\theta_1 - \theta_2}|.
\eeq
With the variance $\Delta\vartheta^2 = 2(\frac{1}{4}\Delta\theta^2)$,
the pair angular distribution $\rho(\vartheta)$ is then approximated as
\beq\label{eq_rho}
\rho(\vartheta) \sim \frac{2}{\sqrt{\pi}\Delta\theta}
\exp\left\{
-\left(\frac{\vartheta}{\Delta\theta}\right)^2
\right\}
\sim \frac{2}{\sqrt{\pi}\Delta\theta}
\quad \mbox{for} \quad 0<\vartheta<\pi/2
\eeq
where
the coefficient 2 of the amplitude is caused by limiting $\vartheta$
to the range $0 < \vartheta < \pi/2$,
and $\left(\frac{\vartheta}{\Delta\theta}\right)^2 \ll 1$
is taken into account because $\Delta\theta$ in Eq.(\ref{eq15})
also corresponds to the upper limit by
the focusing lens based on geometric optics.
This distribution is consistent with the flat top distribution
applied to Ref.\cite{DEapb, DEptep} except the coefficient.

We now re-express the average of the squared scattering amplitude as
a function of $\chi \equiv a\xi$
in units of the width of the Breit-Wigner(BW) distribution $a$
by substituting Eq.(\ref{eq_M2}) and (\ref{eq_rho})
into Eq.(\ref{eq_M2average}) with Eq.(\ref{eq_dvartheta})
\beqa\label{eq_M2a}
\overline{|{\mathcal M}_S|^2}
=
\frac{(4\pi)^2}{\sqrt{\pi} \omega^2_{opt}}
\left(\frac{\vartheta_r}{\Delta\theta}\right) a
{\mathcal W},
\eeqa
where we introduce the following constant
\beq\label{eq_calW}
{\mathcal W} \equiv
\int^{\frac{\omega^2_{opt}}{a} \{1-(\vartheta_r/(\pi/2))^2 \}}_{-\infty}
W(\xi) \frac{1}{\xi^2+1} d\xi
\eeq
with
\beq\label{eq_W}
W(\xi) \equiv (1-\frac{a\xi}{\omega^2_{opt}})^{-3/2}.
\eeq
In Eq.(\ref{eq_calW})
the weight function $W(\xi)$ is the positive and monotonic function within
the integral range and the second term is the Breit-Wigner(BW) function with
the width of unity.
Note that $\overline{|{\mathcal M}_S|^2}$ is now explicitly proportional to $a$
but not $a^2$. This gives the enhancement factor $a$ compared to the case
$\overline{|{\mathcal M}_S|^2} \propto a^2$ where no resonance state is contained
in the integral range controlled by $\Delta\theta$ experimentally.
The integrated value of the pure BW function
from $\xi = -1$ to $\xi = +1$ gives $\pi/2$, while that
from $\xi = -\infty$ to $\xi = +\infty$ gives $\pi$.
The difference is only a factor of two.
The weight function $W(\xi)$ of the kernel is almost unity for small $a\xi$,
that is, when $a$ is small enough with a small mass and a weak coupling.
Therefore, we will consider only the region of $\xi \pm 1$ 
as a conservative estimate. 
By taking only this integral range, 
we can be released from trivial numerical modifications originating
from $\xi =-\infty$ and the behavior of $W(\xi)$ at 
$\xi = \frac{\omega^2_{opt}}{a} \{1-(\vartheta_r/(\pi/2))^2 \}$
which are not essential due to the strong 
suppression by the Breit-Wigner weight.

We now refine $\mathcal{W}$ in order to apply it more accurately even
to the case for $\vartheta_r/\Delta \theta \sim 1$ where, exactly speaking,
the second approximation in Eq.(\ref{eq_rho}) is not valid.
In this case, by using the first of Eq.(\ref{eq_rho}) with substitution
of the relation between $\chi \equiv a\xi$ and $\vartheta$
expressed in Eq.(\ref{exeq_14}), Eq.(\ref{eq_W}) is modified as follows
\beq\label{eq_Wnew}
W(\xi) \equiv \exp\left\{-\frac{(\vartheta_r/\Delta\theta)^2}
{1-\frac{a}{\omega^2_{opt}}\xi}\right\}
(1-\frac{a}{\omega^2_{opt}}\xi)^{-3/2}
\sim \exp\left\{-\left(\frac{\vartheta_r}{\Delta\theta}\right)^2\right\},
\eeq
where the last approximation is based on $a/\omega^2_{opt} \ll 1$
with respect to the integral range $\xi \pm 1$ in Eq.(\ref{eq_calW})
for the conservative estimate.
This is justified in the mass-coupling range we are interested in
via the first relation in Eq.(\ref{eq_a}), for instance,
$a/\omega^2_{opt} \sim 10^{-29}$ 
for $m \sim 0.1$~eV and $g/M \sim 10^{-4}$ GeV${}^{-1}$.
By substituting Eq.(\ref{eq_Wnew}) into Eq.(\ref{eq_calW}),
the conservative evaluation on $\mathcal{W}$ over $\xi \pm 1$ is expressed as
\beq\label{eq_calWnew}
{\mathcal W} \sim
\int^{+1}_{-1}
W(\xi) \frac{1}{\xi^2+1} d\xi
\sim \frac{\pi}{2}
\exp\left\{-\left(\frac{\vartheta_r}{\Delta\theta}\right)^2\right\}.
\eeq
This factor is dependent of $\vartheta_r$, equivalently dependent of mass,
especially for larger $\vartheta_r$ close to $\Delta\theta$ while
it is almost $\pi/2$ for smaller $\vartheta_r$.